\documentclass[12pt,draftclsnofoot, onecolumn]{IEEEtran}
\usepackage{graphicx,cite,epsfig,amssymb,amsmath,subfigure,url,stfloats,latexsym,color}

\begin{document}
\markboth{IEEE Communications Magazine, Feature Topic on Energy
Harvesting Communications, April 2015.} {Chen, Zhang, Chen \& Zhang:
Enhancing Wireless Information \ldots}

\title{Enhancing Wireless Information and Power Transfer by Exploiting Multi-Antenna
Techniques}

\author{\authorblockN{
Xiaoming~Chen, \IEEEmembership{Senior Member, IEEE}, Zhaoyang~Zhang,
\IEEEmembership{Member, IEEE}, \\Hsiao-Hwa~Chen,
\IEEEmembership{Fellow, IEEE}, and Huazi~Zhang,
\IEEEmembership{Member, IEEE}
\thanks{Xiaoming~Chen (e-mail: {\tt chenxiaoming@nuaa.edu.cn}) is
with the College of Electronic and Information Engineering, Nanjing
University of Aeronautics and Astronautics, China. Zhaoyang~Zhang
(e-mail: {\tt ning\_ming@zju.edu.cn}) and Huazi~Zhang (e-mail: {\tt
tom.zju@gmail.com}) are with the Department of Information Science
and Electronic Engineering, Zhejiang University, China.
Hsiao-Hwa~Chen (e-mail: {\tt hshwchen@ieee.org}) is with the
Department of Engineering Science, National Cheng Kung University,
Taiwan.}}}\maketitle

\begin{abstract}
This paper reviews emerging wireless information and power transfer
(WIPT) technique with an emphasis on its performance enhancement
employing multi-antenna techniques. Compared to traditional wireless
information transmission, WIPT faces numerous challenges. First, it
is more susceptible to channel fading and path loss, resulting in a
much shorter power transfer distance. Second, it gives rise to the
issue on how to balance spectral efficiency for information
transmission and energy efficiency for power transfer in order to
obtain an optimal tradeoff. Third, there exists a security issue for
information transmission in order to improve power transfer
efficiency. In this context, multi-antenna techniques, e.g., energy
beamforming, are introduced to solve these problems by exploiting
spatial degree of freedom. This article provides a tutorial on
various aspects of multi-antenna based WIPT techniques, with a focus
on tackling the challenges by parameter optimization and protocol
design. In particular, we investigate the WIPT tradeoffs based on
two typical multi-antenna techniques, namely limited feedback
multi-antenna technique for short-distance transfer and large-scale
multiple-input multiple-output (LS-MIMO, also known as massive MIMO)
technique for long-distance transfer. Finally, simulation results
validate the effectiveness of the proposed schemes.
\end{abstract}

\IEEEpeerreviewmaketitle

\vspace{0.45in}
\section{Introduction}
Wireless power transfer has attracted a lot of attention in wireless
research community, as it can effectively prolong the lifetime of a
power-limited network in a relative simple way, especially under
some extreme conditions, such as battle-field, underwater, and body
areas networks \cite{Medical}. For example, in medical care
applications, devices implanted in body send information to outside
receiver with harvested power from the outside power source.
Recently, wireless power information is proposed for cellular
systems, to provide mobiles practically infinitely long battery
lives and eliminate the need of power cords and chargers. The radio
frequency (RF) signal based wireless power transfer attracts
considerable attention in both academia and industry due to the
following two reasons \cite{SWIPT1} \cite{SWIPT2}. First, it is a
controllable and deterministic power transfer method. For example,
it is possible to flexibly increase transmit power to enhance
receive quality. Second, information and power can be simultaneously
transferred in a form of RF signal. Then, the communications can be
supported without external power sources.

In comparison with conventional wireless information transmission,
wireless information and power transfer (WIPT) exhibits both
similarities and differences. On one hand, both of them suffer
from channel fading and path loss, resulting in performance loss. In
particular, power transfer distance may be relatively short, since
power harvesting is more sensitive than information decoding
\cite{Energybeamforming}. Therefore, it is necessary to effectively
combat the fading effects, so as to improve the efficiency and
distance of power transfer. For traditional wireless information
transmission, multi-antenna technique is a powerful way
to enhance the performance over fading channels. Through spatial
beamforming, multi-antenna techniques can adapt the transmit signal to
channel states, so that channel fading can be harnessed to
improve the performance. Similarly, for wireless power transfer,
multi-antenna technique can also be used to align the RF signal
to a power receiver, thus improving the energy efficiency.
Therefore, it makes sense to exploit the benefits of multi-antenna
technique to enhance the performance of WIPT. On the other hand,
WIPT has two performance metrics, namely spectral efficiency for
information transmission and energy efficiency for power transfer.
In general, the two metrics are inconsistent and even contradictory,
since information and power compete for the same RF signal and
resources. Fortunately, it is convenient for multi-antenna
techniques to achieve a good tradeoff between the spectral and energy
efficiencies by designing appropriate spatial beams for information
and power transfer, respectively \cite{WIPTTradeoff}. More
importantly, multi-antenna techniques may concurrently support multiple
streams of information and power transfer, and thus the efficiencies
are improved significantly.

To exploit the benefits of multi-antenna techniques for WIPT, the
transmitter requires full or partial channel state information
(CSI), and then both information and power are transferred
adaptively to the channel conditions. Specifically, based on the
CSI, a transmitter selects the optimal transmit parameters, i.e.
transmit beam, transmit power, and accessing users in order to
maximize the efficiencies over fading channels. In \cite{FullCSI},
an optimal multiuser WIPT system was designed, assuming that full CSI
is available at the transmitter. However, in multi-antenna systems, it is a
nontrivial task to obtain instantaneous CSI at the transmitter,
since the channel is a multi-dimensional time-varying random matrix.
Generally, according to different duplex modes, there are two CSI
acquisition methods in multi-antenna systems \cite{CSI}. In
frequency division duplex (FDD) systems, the CSI is usually conveyed
from the information and power receivers to the transmitter by
making use of quantization codebooks, so that the transmitter can
obtain partial CSI. Note that a larger codebook size leads to more
accurate CSI, but also increases feedback overheads. Therefore, it
is possible to improve the efficiencies by increasing the feedback
amount. On the other hand, the CSI in time division duplex (TDD) systems
can be estimated at a transmitter, directly making use of channel
reciprocity. Compared to the CSI feedback in FDD systems, CSI
estimation in TDD systems saves the feedback resource, but may
suffer from a performance loss due to transceiver hardware
impairment. To solve the problem, robust beamforming for WIPT was
proposed in \cite{RobustBeamforming} to guarantee high efficiencies
even with imperfect CSI. Moreover, CSI can also be used to construct
transmit beams. However, with respect to the beamforming based on
instantaneous CSI, the one based on estimated CSI suffers an obvious
performance degradation. Thus, adaptive multi-antenna transmission
techniques via CSI feedback or estimation are effective ways to
enhance performance for WIPT over fading channels.

For multi-antenna based WIPT techniques, there are a number of
transmission frameworks proposed in the literature. First, for multi-antenna techniques, there
are several different forms. For example, according to the number
of antennas, there are traditional multi-antenna techniques and
large-scale multiple-input multiple-output (LS-MIMO) techniques.
Additionally, according to the number of accessing users, we have
single-user and multi-user transmission techniques. Second, as
mentioned earlier, there are two different CSI achievement methods,
namely CSI feedback and CSI estimation. Third, according to the
transmission protocols, WIPT can also be classified into two cases.
In the first case, information and power are
transferred simultaneously, namely simultaneous wireless information and power
transfer (SWIPT) \cite{SWIPT}. In the second case, power is first
transferred, and then the harvested power is used to send
information, namely wireless powered communication (WPC) or energy
harvesting communication (EHC) \cite{WPC}. Thus, combining the above
three schemes, multi-antenna based WIPT technique has a variety of
forms, which are applicable to fit to different scenarios. In this article,
we intend to investigate various issues on multi-antenna
based WIPT technique from both theoretical and design perspectives. Especially,
we analyze parameter optimization and protocol design for various
multi-antenna based WIPT techniques. To facilitate understanding, we
use traditional multi-antenna technique and LS-MIMO technique based
WIPTs as two typical examples to instantiate the wireless
information and power transfer tradeoff, and analyze the effect of
CSI accuracy and the number of antennas on the tradeoff.

The rest of this article can be outlined as follows. We give an
introduction of various multi-antenna based WIPT techniques, and then
highlight the parameter optimization and protocol design in Section
II. A discussion and comparison of two typical multi-antenna
techniques for WIPT are given in Section III. Simulation results are
illustrated in Section IV to verify the tradeoff performance of the two
typical multi-antenna based WIPT techniques, followed by the
conclusions and discussions on several open issues in Section V.

\vspace{0.25in}
\section{multi-antenna based WIPT technique}

WPT is not a new technology, although it regains considerable
interests recently. It was developed more than a century ago and its
feasibility has been verified by many practical experiments. At the
end of the 19th century, Nikola Tesla carried out the first WPT
experiment, which tried to transmit approximately 300 KW power via
150 KHz radio waves. In the 1960s, William C. Brown restarted WPT
experiments with high-efficiency microwave technology, and the
efficiency reached to 50$\%$ at an output power of 4W DC. After the
1980s, many experiments were carried out in Japan and the United
States. In the 2000s, advances in microwave technologies pushed WPT
back into consideration for wireless communications. Despite these
advances, there are many challenging issues that remain to be open
for WIPT. It is because that both information and power are carried
by RF signals over wireless media, and they may suffer from
attenuation, noise, interference, and interception. Thus, to
effectively implement WIPT and evaluate the performance, several
fundamental metrics of interest are introduced as follows:
\begin{enumerate}
\item Transfer efficiency: RF signal will decay due to channel
fading caused by reflection, scattering, and refraction in
propagation processes. Thus, the received signal may be very weak, making it
difficult to recover transmit signal or harvest the signal
energy. The problem becomes more prominent for wireless power
transfer, since a power receiver is more sensitive to the
magnitude of RF signal. Hence, it is necessary to improve the
efficiency of WIPT over wireless channels.
\item Transfer distance: The attenuation of RF signal is an
increasing function of transfer distance. To guarantee a viable
received power, wireless power transfer has a stringent limitation
on transfer distance based on the current state of the art research. With an
increasing demand on wireless power transmission, especially
wireless powered communications, this limitation has become a
major bottleneck in the development of wireless power transfer.
Thus, it is imperative to increase the effective transfer distance.
\item Transfer tradeoff: Limited power, spectrum and time resources
are shared by wireless information and power transfer, resulting in
the fundamental tradeoffs between the two. For example, in the
SWIPT, the total transmit power is distributed for information and
power transfer. While in wireless powered communications, each time
slot is divided into information and power transfer durations. To balance
the information and power transfer according to application requirements
and enhance the overall performance, it is vital to analyze the
optimal resource allocation between them.
\item Transfer security: Due to the open nature of wireless media,
information transmission is apt to be overheard. The security issue
is even severer in WIPT because the power receiver is usually placed
closer to the transmitter than the information receiver.
Traditionally encryption technology cannot fully solve the problem,
because it requires a secure channel for exchanging private keys,
becoming impractical in infrastructure-less or mobile networks.
\end{enumerate}

In order to realize efficient, reliable, secure, and long-distance
WIPT, various advanced technologies have been identified recently, such as
cooperative communication, resource allocation, and user scheduling.
In particular, multi-antenna technique has a great potential due to
its significant performance gain. On one hand, multi-antenna
diversity gain can be exploited to combat channel fading as an effort
to improve transfer efficiency and increases transfer distance
\cite{LS-MIMO}. On the other hand, multi-antenna multiplexing gain
can be leveraged to separate information and power transfer in space, so
that the latter two metrics, i.e., transfer tradeoff and transfer
security, can be realized simultaneously \cite{Energybeamforming}. Let us take
a look at a simple example. If the information is transmitted in the null
space of the channel for power transfer, then information security
can be guaranteed with the help of physical layer security,
even the power receiver is very close to the transmitter
\cite{Energybeamforming}. Due to these inherent advantages,
multi-antenna based WIPT technique is receiving a considerable
attention from both academia and industry. In what follows, we give
a detail investigation of multi-antenna WIPT. Due to space
limitation, we only consider single-hop WIPT. In fact, the multi-hop
transmission technology is also a powerful way of enhancing WIPT.
For example, relay technology can shorten the transfer distance,
and thus improve the performance \cite{Relay}. The multi-hop cases
will be studied in the future. According to the transfer model and
protocol, WIPT can be further classified into SWIPT and WPC. Our
investigation will cover a thorough case study of these two models
and their integration.

\subsection{Simultaneous Wireless Information and Power Transfer}
As the name implies, SWIPT transmits information and power
simultaneously. If the transmitter is equipped with multiple
antennas, spatial beamforming adapted to the channel states
can be used to improve the performance of WIPT. In this case, the
information and power receivers can be either combined or separated.
Then, there are two subcases for SWIPT with different design
principles.

\subsubsection{Combined Case}
\begin{figure}[h] \centering
\includegraphics [width=0.8\textwidth] {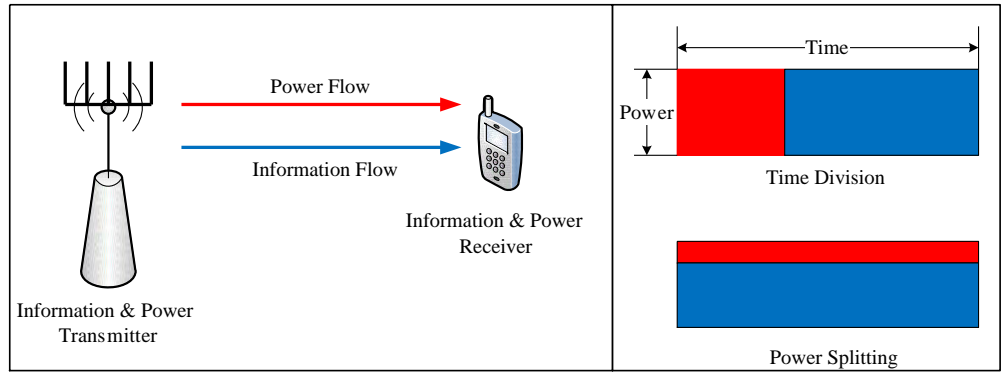}
\caption {Model and protocol for combined case of SWIPT.}
\label{Fig1}
\end{figure}

In this case, a node plays the roles of both information and
power receivers, as shown at the left-hand side of Fig. \ref{Fig1}. The
design of the transmitter is relatively simple. The core step is to
perform spatial beamforming based on the CSI obtained through
feedback in FDD systems or direct estimation in TDD systems.
However, due to the dual roles, the receiver should be designed
carefully. Note that the receiver cannot decode the information and
harvest the energy simultaneously due to physical constraints.
Then, it is required to separate the information and power transfer
by a certain protocol. Currently, there are mainly two protocols,
namely time division protocol \cite{WIPTTradeoff} and power
splitting protocol \cite{PowerSplitting}. Specifically, as shown at
the right-hand side of Fig. \ref{Fig1}, in the time division protocol, each time slot
is divided into information and power transfer durations. Then, the
roles of the receiver should switch between the two. Otherwise, in the power
splitting protocol, the whole received signal is separated into two
parts, one for information decoding and the other for power
harvesting.

Comparing the two protocols, we can find that time division requires
two RF signal receive modules, since the signals for information
decoding and power harvesting are separated at the RF side.
Contrastingly, power splitting only needs one RF signal receiver
module, and the signals for information decoding and power
harvesting are separated at the baseband. Note that there is a
balance or tradeoff between the information transmission and power
transfer, since the time resource for time division and the power
resource for power splitting are constrained and should be allocated
to the two tasks according to a certain optimization objective. For
example, the WIPT tradeoff can be formulated as an optimization
problem of maximizing the information rate subject to a minimum
harvested power or maximizing the harvesting power subject to a
minimum information rate.

Moreover, there may exist multiple receivers in the combined case.
With respect to the single receiver case, there are more challenging
problems to be solved. First, the receivers should be scheduled
according to the urgency of information and power transfer. However,
it is nontrivial to concurently determine the urgency of information and
power transfer. Second, the WIPT tradeoff for each receiver may be
distinct. In other words, each receiver may use different durations
or powers for information decoding. Third, the beam design has
contrasting goals for information and power transfer. For
information transfer, the beams should be designed to mitigate
inter-user interference. However, for power transfer, the inter-user
interference can increase the received power. Fourth, one or more
receivers may be eavesdropper, which gives rise to security
problems. A feasible way for multi-receiver SWIPT is to use time
division multiplexing access (TDMA), such that each time slot is
allocated to only one receiver. Then, the multiple-receiver case is
transformed to multiple single-receiver cases combined with receiver
scheduling. However, the TDMA protocol may be suboptimal with
respect to space division multiplexing access (SDMA) protocol. It is
still an open issue to design an optimal multiple access protocol.

\subsubsection{Separated Case}
\begin{figure}[h] \centering
\includegraphics [width=0.8\textwidth] {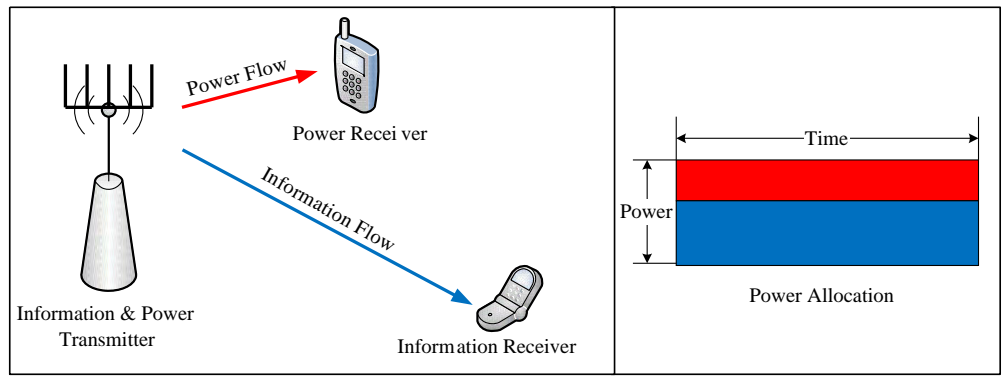}
\caption {Model and protocol for separated case of SWIPT.}
\label{Fig2}
\end{figure}

In the case shown in Fig. \ref{Fig2}, the information and power receivers are separated in
different nodes. The transmitter is allowed to transmit RF
signals for information and power transfer simultaneously in the
same time and frequency resource block. As mentioned earlier, since
the power receiver is more sensitive to the magnitude of RF signal
than the information receiver, it is usually placed closer to the
transmitter, as shown at the left-hand side of Fig. \ref{Fig2}. With
respect to the combined case, the design focus of the separate
case is on the transmitter, but not on the receiver. On one hand,
the transmitter leverages the beamforming to separate the
information and power transfer in space, in order to avoid the
information leakage to the power receiver. On the other hand, the
transmitter needs to allocate the transmit power to two beams,
to achieve a tradeoff between information and power transfer.
For example, the WIPT tradeoff can be formulated as an optimization
problem of maximizing the secrecy rate subject to the minimum
harvested power. It is worth pointing out that, in the sense of
maximizing the secrecy rate, the zero-forcing beamforming (ZFBF) and
the use of maximum transmit power may not be optimal, since ZFBF and
maximum transmit power may reduce the secrecy rate by decreasing the
capacity of the legitimate channel from the transmitter to the
information receiver and increasing information leakage to the
eavesdropper, respectively. If we do not consider the security issues
and only aim to maximize the information rate, the above tradeoff is
reduced to a relatively simple optimization problem.

Similarly, the separated case may also comprise multiple
information and power receivers. If a TDMA or OFDMA protocol is
employed, the problem can be transformed to the case with one
information receiver and multiple power receivers over each time
slot or each subcarrier. In this subcase, if each power receiver, as
an eavesdropper, overhears the information individually, the secrecy
rate is determined by the power receiver with the strongest
interception capability. Otherwise, if the power receivers
cooperatively intercept the information, the secrecy rate is
determined by the combined eavesdropper signal quality. Overall, by
maximizing the sum rate in all slots or subcarriers, it is possible
to get the optimal receiver scheduling and spatial beamforming
schemes. If a SDMA protocol is adopted, all information receivers are
active over the same time-frequency resource block. Then, the
inter-user interference is inevitable, especially with imperfect CSI
at the transmitter. Under such a circumstance, the design of transmit
beams is more complicated, and is still an open issue.

\subsection{Wireless Powered Communication}

\begin{figure}[h] \centering
\includegraphics [width=0.8\textwidth] {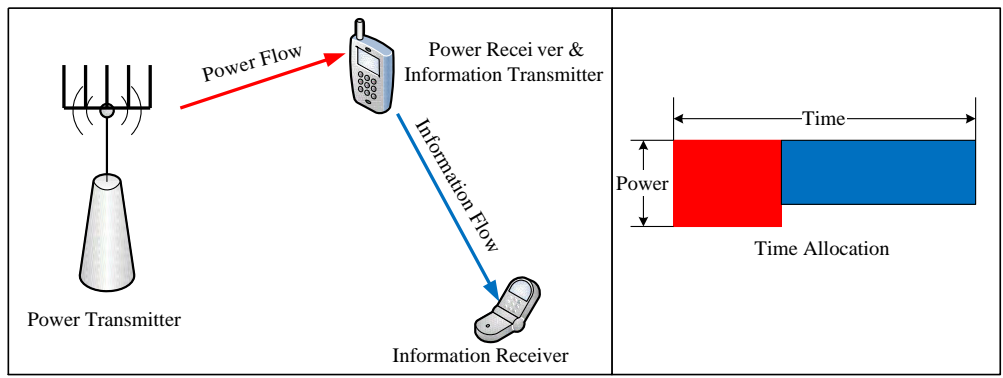}
\caption {Models and protocols for WPC.} \label{Fig3}
\end{figure}

Different from SWIPT, WPC uses the harvested power to transmit
information, and thus it is also named energy harvesting communications. As
a simple example, in medical care applications, the implanted equipment
transmits the information it monitors to the instrument outside
with the harvested power, as seen at the left-hand side of Fig.
\ref{Fig3}. With respect to SWIPT, WPC combines information and
power transfer more closely, since the harvested power may also
affect the information rate.

For the design of WPC, it is important to achieve the optimal
tradeoff between information and power transfers. For example, based
on the time division protocol, the tradeoff is to determine a
switching point between power and information transfers, as shown at
the right-hand side of Fig. \ref{Fig3}. Since the power for information
transmission comes solely from energy harvesting, the tradeoff based
on the time division protocol can be formulated as an optimization
problem maximizing the information rate with a given transmit power
or minimizing the transmit power subject to a minimum rate.

More recently, several new technologies were introduced to further
enhance the performance of multi-antenna based WIPT techniques. For
example, large-scale MIMO technique can generate high-resolution
spatial beams by deploying tens or even hundreds antennas. The
benefit of large-scale MIMO technology for WPC lies in two-fold
\cite{LS-MIMO}. First, the transfer efficiency and distance can be
significantly improved by making use of its large array gain, so as
to enable long-distance WPC with low power. Second, the
high-resolution beam can reduce the information leakage to an
unintended node to achieve information security. As the
number of antennas increases, the performance gain becomes larger,
which is a main advantage of LS-MIMO technique based WPC.

\subsection{Integration of SWIPT and WPC}

In fact, SWIPT and WPC can be integrated to give a more general WIPT
scenario described as follows. First, the transmitter sends
information and power to one or multiple receivers, and then the
power receivers send information to their next-hop receivers using
the harvested power. The design of such a general WIPT can be
considered as a concatenation of SWIPT and WPC. Its transmission
protocol is also based on an integration of SWIPT and WPC
components. In other words, each time slot is divided into two
durations, one for SWIPT and the other for WPC.

If the time division protocol is adopted at SWIPT stage, each time
slot is partitioned into three non-overlapped durations. Typically,
the power receiver will allocate constrained time duration to
either receiving information from the power transmitter, or
transmitting information to the next-hop receiver, which may lead to a
low efficiency. Actually, it is possible to transmit and receive
information simultaneously at the power receiver with recently
introduced full-duplex technology \cite{Full-Duplex}. For example,
if the information transmitter at SWIPT stage is also the
information receiver at WPC stage, the current full-duplex
technology can be exploited to improve the efficiency. A potential
problem of the full-duplex technology is the self interference from the
information transmitter to the receiver. Fortunately,
multi-antenna technology can be used to cancel the self-interference
by making use of the spatial degrees of freedom. Hence,
multi-antenna based WIPT technique combining full-duplex can
significantly improve the performance.

\begin{figure}[h] \centering
\includegraphics [width=0.8\textwidth] {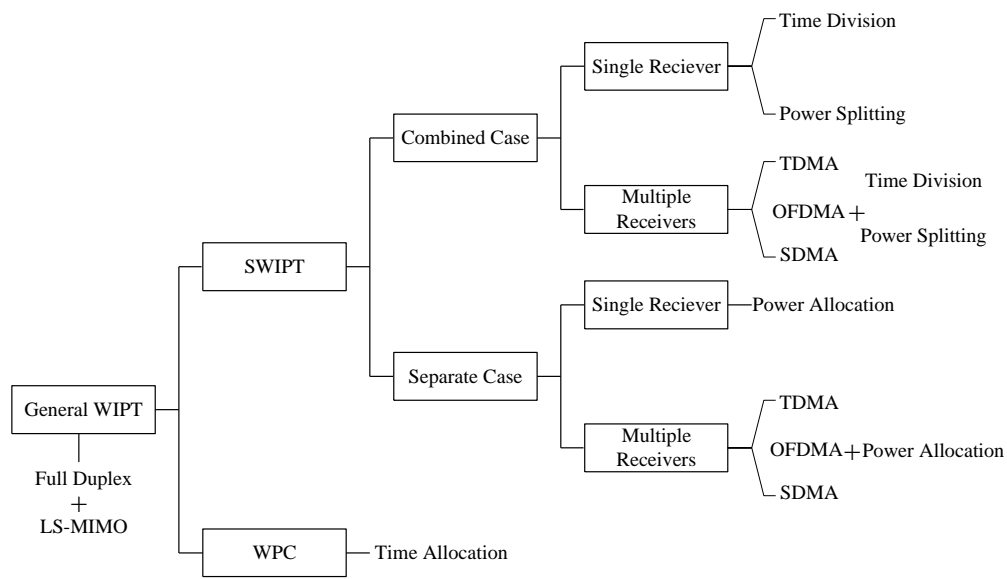}
\caption {A summary of multi-antenna based WIPT.} \label{Fig4}
\end{figure}

In all scenarios, multi-antenna based WIPT technique can solve a
series of challenging issues, making it an attractive solution to
provide efficient, reliable, secure, and long-distance transfer. In
Fig. \ref{Fig4}, we give a summary of various multi-antenna based
WIPT techniques together with their corresponding transfer
protocols.

\vspace{0.25in}
\section{Wireless Information and Power Transfer Tradeoff}

In this section, we focus on the tradeoff or balance between
wireless information and power transfer in single user multi-antenna
systems. As discussed earlier, information and power transfers have
different performance metrics. For example, information transmission
mainly concerns the rate, delay, and security, while power transfer
emphasizes the efficiency and distance. Intuitively, the goals for
information and power transfers are inconsistent, and even
contradictory. Thus, it is of importance to achieve an optimal
tradeoff in the design of WIPT.

It is a common practice to achieve the performance objectives by
optimizing the system parameters, e.g., transmit beam, transmit
power, transfer duration, user scheduling, channel selection, and
transfer protocol. For SWIPT, since the performance objectives are
relatively independent, the tradeoff is usually formulated as three
types of optimization problems. First, a multi-objective
optimization scheme can be adopted, in order to maximize the two
performance indexes simultaneously. Second, the objective can be
expressed as a general utility function. For example, it is
reasonable to take a weighted sum of the efficiency as the
objective. Third, the problem can be formulated by maximizing one
performance index subject to a constraint on the other performance
indices. For instance, a common problem in the existing related literatures
is to maximize the information rate subject to a minimum harvesting
power constraint. Different from SWIPT, WPC relates the two
performance metrics more closely, since the harvested power is used
for information transmission. Therefore, the tradeoff for WPC has a
direct and single formulation. In what follows, through two typical
tradeoffs for multi-antenna technique based WPC, we present their
protocol designs and parameter optimizations.

\subsection{Information Rate Maximization in Traditional Multi-Antenna Systems}
First, let us consider a traditional multi-antenna based WPC technique,
as shown in Fig. \ref{Fig3}. A multi-antenna power transmitter
charges a power receiver via RF signals at the beginning of each
time slot, and then the power receiver sends information to an
information receiver. Note that the power transmitter and the
information receiver can be the same node in some cases. This is a
typical application scenario in medial care (e.g., microchip
implant) and underwater monitoring.

In order to improve the power transfer efficiency and thus maximize
the information transmission rate, energy beamforming is conducted
at the power transmitter. In practice, the multi-antenna power
transmitter directs the RF signals to the receiver according to the
current channel state, so as to overcome the negative effects of
channel fading and propagation loss. Note that the performance of
energy beamforming depends on the accuracy of CSI at the
transmitter. As mentioned earlier, the CSI is obtained through
feedback in FDD systems or direct estimation in TDD systems.
Considering the fact that the power transmitter and the information
receiver are in general separate, limited feedback based on a
quantization codebook is a more practical choice. In such a system,
the harvested power at the information transmitter can be considered
as an increasing function of CSI feedback amount, transfer duration,
and transmit power, and at the same time a decreasing function of
transfer distance.

With the harvested power, the information transmitter sends
information to the receiver in the remaining time of the slot. In
general, the average transmit power for information transmission is
equal to the quotient of the harvest energy and the duration left
for information transmission. Therefore, according to Shannon
capacity equation, the average amount of information transmitted
during a time slot can be expressed as a function of the average
transmit power. Finally, the average information transmission rate
can be derived through dividing the average amount of information
transmitted during a time slot by the length of a time slot.
Intuitively, it is a function of transmit power at the power
transmitter, power transfer duration, and CSI feedback amount.

Taking the maximization of average information transmission rate as
the optimization objective, we can derive the optimal transfer
duration for a given CSI feedback amount, namely the switching point
for power and information transfers. By adjusting the amount for CSI
feedback, we can get different tradeoffs.

So far, we have only given a basic example. In fact, it can be
extended to several more complex cases. First, when the information
transmission has a certain quality of service (QoS) requirement, the
above optimization problem should include a QoS constraint. It is
worth pointing out that, given transmit power and feedback amount,
there may be no feasible solutions for transfer duration. To solve
it, we should increase transmit power or feedback amount.
Additionally, if the system is power-limited, we can formulate the
problem as minimizing the transmit power, while satisfying the QoS
requirement. Second, the basic model can also be naturally extended
to the case of a general WIPT. Similarly, we need to add a minimum
rate constraint for the information transmission from the power
transmitter to the power receiver. Meanwhile, if time division
protocol is adopted, an optimization variable of information
transfer duration should be added. Otherwise, if power splitting
protocol is adopted, the added optimization variable should be the
power splitting ratio instead. Third, in the case of an eavesdropper
overhearing the information sent from the information transmitter,
the above optimization problem is transformed to maximizing the
secrecy rate.

\subsection{Energy Efficiency Maximization in LS-MIMO Systems}
LS-MIMO technique can generate a high-resolution spatial beam through
the deployment of a large number of antenna elements, and thus
achieve substantial transfer efficiency and distance gains. In this
case, we consider a WPC system, where both the power transmitter and the
information receiver are equipped with a large-scale antenna array.

To fully exploit the benefits of LS-MIMO techniques, the transmitter
needs to know the exact CSI. However, due to a large amount of
feedback (proportional to the number of antennas) in LS-MIMO
systems, the CSI feedback scheme is practically infeasible. Thus,
LS-MIMO systems usually work in TDD mode, and therefore the CSI can
be estimated by making use of channel reciprocity. However, due to
transceiver hardware impairment, the estimated CSI may be imperfect,
resulting in certain performance loss. Hence, the CSI
accuracy is also a decisive factor in determining the performance.
Note that the numbers of antennas at the power transmitter and the
information receiver are usually quite large (e.g., more than 100).
According to the law of energy conservation, the harvested power at
the power receiver (namely the information transmitter) is a
function of transmit power at the power transmitter, power transfer
duration, and CSI accuracy based on TDD mode. In addition, due to
channel hardening in LS-MIMO system, it is also a deterministic
function of the number of transmit antennas. Similarly, with the
harvested power, the average information transmission rate can be
expressed as a function of transmit power, transfer duration, CSI
accuracy, the number of power transmit antennas, and the number of
information receiver antennas by making use of Shannon capacity
expression.

The energy efficiency, defined as the bits transferred per Joule
energy, is a key performance metric for WIPT
\cite{EnergyEfficiency}. Therefore, we maximize the energy
efficiency to get the optimal tradeoff for such an LS-MIMO
based WPC system. As pointed out earlier, the amount of information
transferred during a time slot can be computed through multiplying
the average information transmission rate by the length of a time
slot, and the total energy consumption is the sum of the energy
consumption in the power amplifier at the power transmitter and the
constant energy consumption in the transmit filter, mixer, frequency
synthesizer, and digital-to-analog converter (which are independent
of the actual transmit power). Hence, by maximizing the ratio of the
amount of information transmission and the total energy consumption,
we can derive the optimal transfer duration. Similarly, we can add
QoS and secrecy requirements on the basis of the above problem. Note
that if the extended problem has no feasible solutions, we can make
it feasible by simply adding more antennas at the power transmitter
or the information receiver, which is a main advantage of the
LS-MIMO based WIPT techniques.

\vspace{0.25in}
\section{Performance Analysis and Simulations}
In this section, we present some simulation results to validate the
tradeoffs of multi-antenna based WPC technique, where the power
transmitter and the information receiver are integrated in one node.
The parameters used are defined as follows. We set the length of a time slot as
$T=5$ ms, noise variance $\sigma^2=-125$ dBm, energy conversion
efficiency from RF signals to electric energy $\theta=0.9$, constant
power consumption $P_0=30$ dBm, and path loss for power transfer and
information transmission $\alpha=\beta=10^{-2}d^{-\nu}$, where $d$
is the transfer distance and $\nu=4$ is the path loss exponent. Note
that, in the given path loss model, a path loss of 20 dB is assumed
at a reference distance of 1 meter. In addition, we use $B$ and $\rho$
to denote the feedback amount in traditional multi-antenna systems
and the CSI accuracy in LS-MIMO systems, respectively.

\begin{figure}[h] \centering
\includegraphics [width=0.8\textwidth] {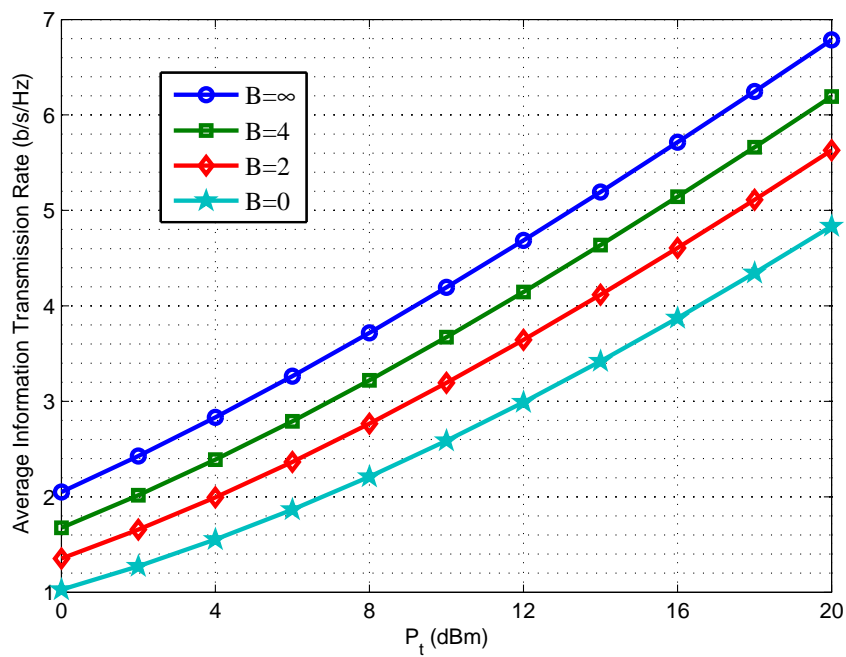}
\caption {Information rate of traditional multi-antenna based WPC
technique with different feedback amounts.} \label{Fig5}
\end{figure}

First, let us consider the tradeoff for a traditional multi-antenna
based WPC technique with $N_t=N_r=4$ and $d=10$ m. As discussed in
Section III.A, we take the maximization of average information
transmission rate as the optimization objective and adjust the
transfer duration. It is shown in Fig. \ref{Fig5} that the feedback
amount $B$ has a great impact on the tradeoff, and thus affects the
information rate. In comparison to the case without feedback, a
small feedback amount, e.g., $B=2$, can increase the information
rate remarkably. However, as the amount of feedback increases, the
additional gain in term of information rate diminishes. As seen from
the results, with a finite feedback amount of $B=4$, the performance
gap to the ideal case (full feedback) is small. Thus, the insight
obtained here is that, with even limited CSI feedback, the
traditional multi-antenna technique can effectively enhance the
performance of WIPT.

\begin{figure}[h] \centering
\includegraphics [width=0.8\textwidth] {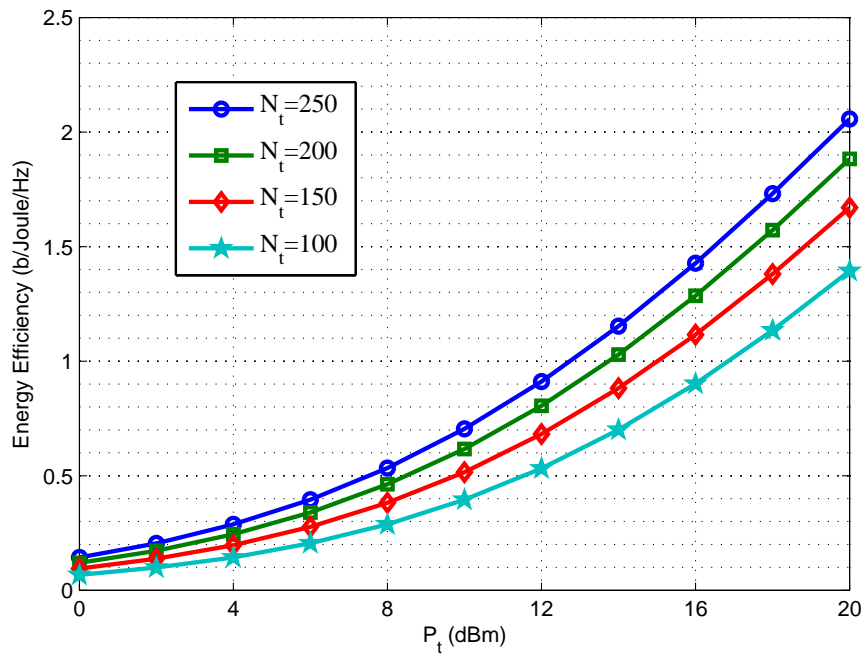}
\caption {Energy efficiency of LS-MIMO based WPC technique with
different numbers of antennas.} \label{Fig6}
\end{figure}

Second, let us examine the effect of LS-MIMO technique on the
tradeoff of WPC with $N_r=100$, $\rho=0.9$, and $d=50$ m. This
corresponds to a long-distance power transfer scenario. With respect
to traditional multi-antenna techniques, only LS-MIMO technique can
support such a long transfer distance without consuming more
transmit power, which is a very appealing characteristic feature.
Take energy efficiency as the optimization metric, we derive the
optimal tradeoff of LS-MIMO technique based WPC, as shown in Fig.
\ref{Fig6}. It is found that the number of antennas has a great
impact on the energy efficiency, which validates an antenna number
versus energy efficiency tradeoff. By adding more antennas, the
energy efficiency can be improved further, which enables a high QoS
for WPC with an affordable power even in the presence of imperfect
CSI.

\vspace{0.125in}
\section{Conclusion and Future Works}

This article reviewed the key technologies in WIPT and discussed
several challenging issues, i.e., transfer efficiency, distance,
tradeoff, and security. Through summarizing the existing works on
multi-antenna based WIPT techniques, this paper gives a comprehensive
tutorial covering both parameter optimization and protocol design,
and proposes to use full-duplex and LS-MIMO technologies to
solve the challenges in various WIPT scenarios. In particular,
a concept of WIPT tradeoff based on multi-antenna
technique is introduced and analyzed in detail. Finally, the
tradeoffs are validated through simulations using the proposed
schemes in two typical multi-antenna scenarios.

It is worth pointing out that there are still many open issues for
WIPT, especially for multiuser WIPT. First, the user scheduling
schemes should be carefully designed to balance the QoS
requirements, resource constraints, and information security. Second,
transmit beams need to be elaborately constructed to achieve a
proper tradeoff between information and power transfers, in
particular with imperfect CSI. Third, the benefits of advanced
multi-antenna techniques for WIPT should be further exploited. For
instance, the self-interference of full-duplex techniques is adverse
to information transmission, but can be harnessed to enhance power
transfer. Hence, it is not optimal to cancel the self-interference
completely, and a more in-depth investigation is required.


\vspace{0.2in}
\begin{thebibliography}{1}
\bibitem{Medical}
F. Zhang, S. A. Hackworth, X. Liu, H. Chen, R. J. Sclabassi, and M.
Sun, ``Wireless energy transfer platform for medical sensor and
implantable devices," in \emph{Proc. IEEE EMBS 31st Annual Int.
Conf.}, pp. 1045-1048, Sept. 2009.

\bibitem{SWIPT1}
P. Grover, and A. Sahai, ``Shannon meets Tesla: wireless information
and power transfer," in \emph{Proc. IEEE Int. Symp. Inf. Theory
(ISIT)}, pp. 2363-2367, June 2010.

\bibitem{SWIPT2}
K. Huang, and E. G. Larsson, ``Simultaneous information and power
transfer for broadband wireless systems," \emph{IEEE Trans. Signal
Process.}, vol. 61, no. 23, pp. 5972-5986, Dec. 2013.

\bibitem{Energybeamforming}
L. Liu, R. Zhang, and K-C. Chua, ``Secrecy wireless information and
power transfer with MISO beamforming," \emph{IEEE Trans. Signal
Process.}, vol. 62, no. 7, pp. 1850-1863, Apr. 2014.

\bibitem{WIPTTradeoff}
X. Chen, C. Yuen, and Z. Zhang, ``Wireless energy and information
transfer tradeoff for limited feedback multiantenna systems with
energy beamforming," \emph{IEEE Trans. Veh. Technol.}, vol. 63, no.
1, pp. 407-412, Jan. 2014.

\bibitem{FullCSI}
R. Zhang, and C. K. Ho, ``MIMO broadcasting for simultaneous
wireless information and power transfer," \emph{IEEE Trans. Wirless
Commun.}, vol. 12, no. 5, pp. 3543-3553, May 2013.

\bibitem{CSI}
D. J. Love, R. W. Heath, V. N. Lau, D. Gesbert, R. D. Rao, and M.
Andrews, ``An overview of limited feedback in wireless communication
systems," \emph{IEEE J. Sel. Areas Commun.}, vol. 26, no. 8, pp.
1341-1365, Oct. 2008.

\bibitem{RobustBeamforming}
Z. Xiang, and M. Tao, ``Robust beamforming for wireless information
and power transmission," \emph{IEEE Wireless Commun. Lett.}, vol. 1,
no. 4, pp. 372-375, Aug. 2012.

\bibitem{SWIPT}
I. Krididis, ``Simultaneous information and energy transfer in
large-scale networks with/without relaying," \emph{IEEE Trans.
Commun.}, vol. 62, no. 3, pp. 900-912, Mar. 2014.

\bibitem{WPC}
X. Lu, P. Wang, D. Niyato, D. I. Kim, and Z. Han, ``Wireless
networks with RF energy harvesting, a contemporary survey,"
\emph{IEEE Commun. Surveys \& Tutorials}, 2015. [Online]:
http://arxiv.org/abs/1406.6470v1.

\bibitem{PowerSplitting}
L. Liu, R. Zhang, and K-C. Chua, ``Wireless information and power
transfer: a dynamic power splitting approach," \emph{IEEE Trans.
Commun.}, vol. 61, no. 9, pp. 3990-4001, Sept. 2013.

\bibitem{LS-MIMO}
X. Chen, X. Wang, and X. Chen, ``Energy-efficient optimization for
wireless information and power transfer in large-scale MIMO systems
employing energy beamforming," \emph{IEEE Wireless Commun. Lett.},
vol. 2, no. 6, pp. 667-670, Dec. 2013.

\bibitem{Relay}
D. S. Mishalogoulos, H. A. Suraweera, and R. Schober, ``Relay
selection for simultaneous information transmission and wireless
energy transfer: a tradeoff perspective," \emph{IEEE J. Sel. Areas
Commun.}, vol. 33, Feb. 2015. [Online]:
http://arxiv.org/abs/1303.1647v1.

\bibitem{Full-Duplex}
C. Zhong, H. A. Suraweera, G. Zheng, I. Krididis, and Z. Zhang,
``Wireless information and power transfer with full duplex
relaying," \emph{IEEE Trans. Commun.}, vol. 62, no. 10, pp.
3447-3461, Oct. 2014.

\bibitem{EnergyEfficiency}
D. W. K. Ng, E. S. Lo, and R. Schober, ``Energy-efficient resource
allocation in OFDMA systems with hybrid energy harvesting base
station," \emph{IEEE Trans. Wireless Commun.}, vol. 12, no. 7, pp.
3412-3427, Jul. 2013.

\end{thebibliography}
\end{document}